\documentclass[doublecol]{epl2}
\usepackage{amsmath} 

\title{Effects of phase transitions in devices actuated by the electromagnetic vacuum force}

\author{A. Benassi\inst{1,2} \and C. Calandra\inst{2}}
\shortauthor{A. Benassi \and C. Calandra}

\institute{                    
  \inst{1} 
CNR/INFM-National Research Center on nanoStructures and bioSystems at Surfaces (S3)\\
Via Campi 213/A, I-41100 Modena, Italy \\
  \inst{2} 
Dipartimento di Fisica, Universit\'a di Modena e Reggio Emilia\\
Via Campi 213/A, I-41100 Modena, Italy}


\abstract{We study the influence of the electromagnetic vacuum force on the behaviour of a model device based on materials, like germanium tellurides, that undergo fast and reversible metal-insulator transitions on passing from the crystalline to the amorphous phase.  The calculations are performed at finite temperature and fully accounting for the behaviour of the material dielectric functions. The results show that the transition can be exploited to extend the distance and energy ranges under which the device can be operated without undergoing stiction phenomena. We discuss the approximation involved in adopting the Casimir expression in simulating nano- and micro- devices at finite temperature.}

\begin{document}

\maketitle 
\section{Introduction}
With the development of modern technology towards smaller structures an increasing attention has been addressed to the role of 
electromagnetic vacuum fluctuation forces (dispersion or van der Waals and Casimir forces) in micro- and nano-electromechanical systems 
\cite{serry,serry2,buks2,buks,santos,delrio,batra}. These forces vary 
typically with the third or fourth inverse power of the distance between the surfaces of the interacting bodies and therefore can be very intense in 
the sub-micrometer regime. When the moving parts of a device come to such a close distance, dispersion forces affect the dynamics and the operation 
of the device and, under certain circumstances, they can cause adhesion or stiction between the surfaces, thus limiting the device lifetime. 
Devices actuated by Casimir force have been designed and demonstrated \cite{chan1,chan2,ashourvan}.
Since these forces depend upon the reflectivity, the size and the 
geometry of the interacting parts \cite{Iannuzzi,chen2}, several possibilities can be explored to devise structures that exploit such dependence in order to improve the 
performance of existing devices or to design new types of devices \cite{barcenas}.\\
In this paper we investigate the possibility of using materials, that undergo fast and reversible phase transitions, to extend the travel range of a 
device. Thermally induced phase changes between the amorphous and the polycrystalline state in a thin film have been observed since a long time in 
tellurides containing Ge and Sb (GST materials) \cite{ovshinsky,chopra,bahl1,bahl2,yamada,yamada2,libera,lee}. The transition is accompanied by 
significant changes in optical and transport properties, a feature that is exploited in optical data storage \cite{Ohta} and could be useful in new 
nanoscale memories \cite{lankhorst,wuttig}. 
\begin{figure}
\centering
\includegraphics[width=8cm,angle=0]{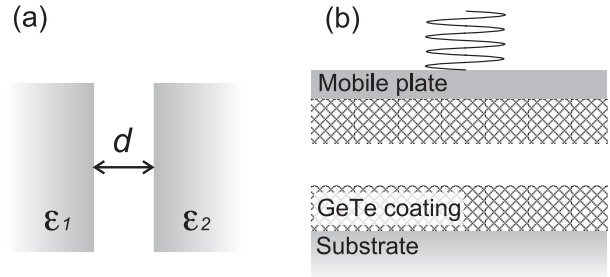}
\caption{\label{fig1}(a) Three layers model of equation
(\ref{force}); (b) Simple representation of the dispersion forces
based device.}
\end{figure}
Previous theoretical studies have shown that metal insulator transitions may cause significant changes in the Casimir force behaviour \cite{pirozhenko2,esquivel2}.
In this paper we propose to use this transition to modify the force between the components of a simple device. 
We show that the locations of the extrema in the total energy potential 
curve of the device are displaced by the transition in such a way that the range of distances and the energy interval over which the device can be 
operated can be significantly modified.
\section{The model device}
We consider an ideal actuator \cite{serry,barcenas} consisting of two parallel plates separated by a gap, with one plate fixed on a substrate and the other suspended by 
an elastic restoring force $F=-K x_0 \delta$, where $x_0$ is the unactuated distance and $K$ is the spring constant. The stationary plate has a flat 
surface at $\delta=1$, while $\delta=0$ denotes the equilibrium position of the movable plate in the absence of dispersion forces and corresponds to 
the unstretched state of the spring. The attractive interaction between the plates of Fig. \ref{fig1}(a) is provided by the electromagnetic vacuum 
fluctuation force, whose expression in terms of the temperature $T$, the dielectric functions $\epsilon_{1}$ and $\epsilon_{2}$ and the 
inter-plate distance $d=(1-\delta)x_0$, is given by the Lifshitz formula \cite{lifshitz,dzyaloshinskii}:
\begin{equation}
\begin{split}
F=&-\frac{A}{\pi\beta}\sum_{n=0}^{\infty \prime} \int_{0}^{\infty}\gamma k dk\bigg[
\frac{1-Q_{TM}(i\Omega_{n})}{Q_{TM} (i\Omega_{n})}+\\
+&\frac{1-Q_{TE}(i\Omega_{n})}{Q_{TE}(i\Omega_{n})} \bigg]
\end{split}
\label{force}
\end{equation}
here $A$ is the plate surface, $\beta=1/k_{B}T$, $k_{B}$ is the Boltzmann constant,
$\Omega_{n}=2 \pi n / \hbar \beta$ is the Matsubara frequency
corresponding to the $n$-th thermal fluctuation mode, the prime on
the summation indicates that the $n=0$ term is given half weight.
We assume the plate size to be much larger than the separation distance.
$Q_{TM}$ and $Q_{TE}$ refer to transverse magnetic (TM) and
transverse electric (TE) modes respectively and are given by:
\begin{subequations}
\begin{align}
Q_{TM}(i\Omega_n)=1&-\frac{(\epsilon_{1}\gamma-\gamma_{1})
(\epsilon_{2}\gamma-\gamma_{2})}{(\epsilon_{1}\gamma+\gamma_{1})
(\epsilon_{2}\gamma+\gamma_{2})}e^{-2d\gamma}\\
Q_{TE}(i\Omega_n)&=1-\frac{(\gamma-\gamma_{1})
(\gamma-\gamma_{2})}{(\gamma+\gamma_{1})
(\gamma+\gamma_{2})}e^{-2d\gamma}
\end{align}
\end{subequations}
with:
\begin{equation}
\gamma_{i}^2=k^{2}+\frac{\Omega_{n}^{2}}{c^{2}}\epsilon_{i}(i\Omega_{n})
\qquad \gamma^2=k^{2}+\frac{\Omega_{n}^{2}}{c^{2}} 
\label{gamma}
\end{equation}
and the dielectric functions are evaluated at the frequency $i\Omega_{n}$ by means of the London transform:
\begin{equation}
\epsilon(i\omega)=1+\frac{2}{\pi}\int_{0}^{\infty}\frac{\Im [ \epsilon(y)]}{y^2+\omega^2}\:y\:dy
\end{equation}
where $\Im [ \epsilon(y)]$ is the imaginary part of the dielectric function.
In the calculation for the metallic phase we follow the prescription of ref. \cite{milton}, i.e. we put the $n=0$ contribution of the TE modes equal zero.
The behaviour of the force as a function of $d$ is non linear: at short distances it can be reproduced by a third inverse power of $d$ (van 
der Waals regime), but a single inverse power term does not reproduce its behaviour for $d$ values larger than a few tens of nanometers. As 
pointed out by several authors \cite{ekinci,zhang}, mainly in connection with electrostatic actuators, there is an intrinsic instability in a device of this sort, which 
prevents the plates to be stably positioned over a large distance. The system is bistable, having a total potential energy with a local and an 
absolute minimum separated by a barrier. This limits the range of motion of the device and determines stiction at short distances between the plates 
\cite{serry,serry2,buks,santos,delrio}. 
We adopt the configuration illustrated in Fig. \ref{fig1}(b) where both plates are made of telluride films deposited on proper substrates. For simplicity we 
assume that the thickness of the deposited films is larger than the distance between the plates (if the film size is comparable to the distance, the 
substrate is expected to affect the intensity of the force). This assumption can be removed by extending the expression of the Lifshitz force to deal 
with a five layer system \cite{zhou,barcenas}.
\section{GeTe dielectric properties and the force}
\begin{figure}
\centering
\includegraphics[width=8cm,angle=0]{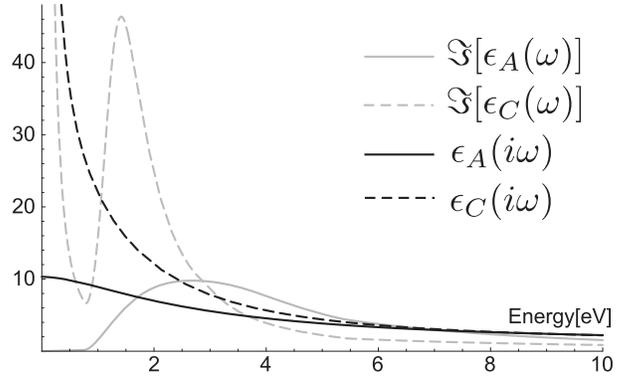}
\caption{\label{fig2}Imaginary parts of amorphous and crystalline
GeTe dielectric functions and their London transforms, adapted from ref.  \cite{welnic} as discussed in the text.}
\end{figure}

To illustrate our model device, we focus our attention on GeTe, whose optical properties have been experimentally determined in both phases. It 
undergoes rapid transitions between polycrystalline and amorphous states under either optical or electrical excitations. According to the band theory 
in a perfect crystal rocksalt structure GeTe should exhibit a semiconducting behaviour. However the stable polycrystalline phases are 
characterized by large vacancy concentrations (typically $10^{20}$ $cm^{-3}$) and local distortions \cite{welnic2,wuttig2}. Compared to other defects, that may be 
present, germanium vacancies have the lower formation energy \cite{edwards}. GeTe thin films show a $p$-type electrical conductivity with a high carrier 
density of up to 
$5\times 10^{20}$ $cm^{-3}$. This density is consistent with the concentration of Ge vacant sites, i.e. the metallic conductivity is the 
consequence of the holes that are formed per single vacant site. In the optical properties direct evidence of the metallic behaviour is provided by 
the presence of Drude-like intraband absorption in crystalline phase \cite{welnic} (see Fig. \ref{fig2}). 
On the other hand the amorphous phase has typical 
semiconductor properties, as illustrated by the behaviour of the imaginary part of the dielectric function, reported in the same figure. Conventional 
semiconductors, such as Si or GaAs, do not show such large differences in the optical properties as a consequence of the amorphous-crystal 
transition.\\
In Fig. \ref{fig2} we display the calculated London transform for both phases. To obtain this curve we have reproduced the experimental data available in 
the range of frequencies $0.1\div 5$ $eV$ with a Drude-Lorentz model including both intra- and inter-band transitions in the metallic phase, and pure inter-band 
transitions in the amorphous phase. The values of the carrier density, the carrier effective mass, the relaxation time and the plasma 
frequency agree with those derived from various experimental studies \cite{chopra,bahl1,bahl2}. 
It is seen that $\epsilon(i\omega)$ turns out to be very different in the two phases in almost all the range of frequencies. As a consequence of the intraband 
absorption, that is present in the metallic phase only, the curve for the polycrystalline phase for small $\omega$ takes much higher values than the 
one for the amorphous phase. Such differences over all the integration interval of equation (\ref{force}) are expected to determine 
significant changes in the force. 
We can use equation (\ref{force}) to calculate the force between two amorphous coated plates $F_{AA}$, two crystal coated plates $F_{CC}$, and the mixed configuration $F_{CA}$.
\begin{figure}
\centering
\includegraphics[width=8cm,angle=0]{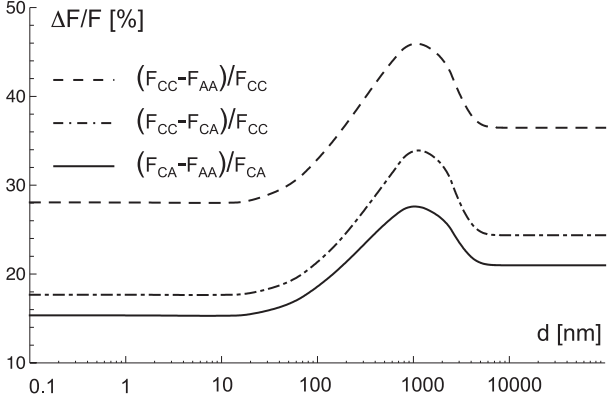}
\caption{\label{fig3}Relative difference between $F_{CC}$,
$F_{CA}$ and $F_{AA}$.}
\end{figure}
In Fig. \ref{fig3} we report, as a function of the separation distance, the relative variation of the dispersion force between the plates with respect to the 
configuration with both plates in the crystalline phase, when one or both plates undergo the transition from the crystalline to the amorphous phase. 
The curves have a similar behaviour with two ranges, at small and large distances, where the relative difference is constant, and a maximum in the range 
between $50$ $nm$ and $5$ $\mu m$. The constant values are a consequence of the fact that the forces have identical behaviour as a function of the plate separation. At short distances the force can be approximated by \cite{hartmann}:
\begin{equation}
F=-\frac{1}{4 \pi \beta d^3}\sum_{n=0}^{\infty \prime}\sum_{m=1}^{\infty}\frac{1}{m^3}\bigg(\frac{\epsilon_{1}-1}{\epsilon_{1}+1}\bigg)^m\bigg(\frac{\epsilon_{2}-1}{\epsilon_{2}+1}\bigg)^m
\label{ecco1}
\end{equation}
which gives a relative change of approximately $28\%$, that does not depend upon the distance and the temperature.
Notice that for the $n=0$ term of the Matsubara sum, the crystalline dielectric function diverges and:
\begin{equation}
\frac{\epsilon-1}{\epsilon+1}\rightarrow 1 \qquad \sum_{m=1}^{\infty}\frac{1}{m^3}=\zeta(3)
\end{equation}
where $\zeta(x)$ is the Riemann zeta function.
On the other hand, at large distances within the present treatment of the temperature dependence of the Casimir force, the force between polycrystalline plates is simply given by \cite{milton,hoye}\footnote{Notice that this behaviour is different from the expression of the Casimir force at $T=0^{\circ}$ $K$ that for a perfect metal gives $F_{CC}=-\frac{\hbar c \pi^2}{240 d^4}$. This formula has been frequently used in previous studies of the role of electromagnetic forces in stiction phenomena.}:
\begin{equation}
F_{CC}\simeq -\frac{\zeta(3)}{8 \pi \beta d^3}
\label{ecco2}
\end{equation}
In the amorphous case the force for large $d$ cannot be written in a closed form. However it can be very well approximated by the $n=0$ term of the Matsubara sum 
\begin{equation}
F_{AA}\simeq -\frac{1}{8 \pi \beta d^3} Li_{3}\bigg[\bigg(\frac{\epsilon_{A}(0)-1}{\epsilon_{A}(0)+1}\bigg)^2\bigg]
\end{equation}
where $Li_{n}[x]$ is the polylogaritmic function of $n$-th order in the argument $x$.
From these expressions one obtains a relative change of the order of $36\%$, in agreement with the exact numerical results in Fig. \ref{fig2}.
\section{Tailoring the device performance}
\begin{figure}
\centering
\includegraphics[width=8cm,angle=0]{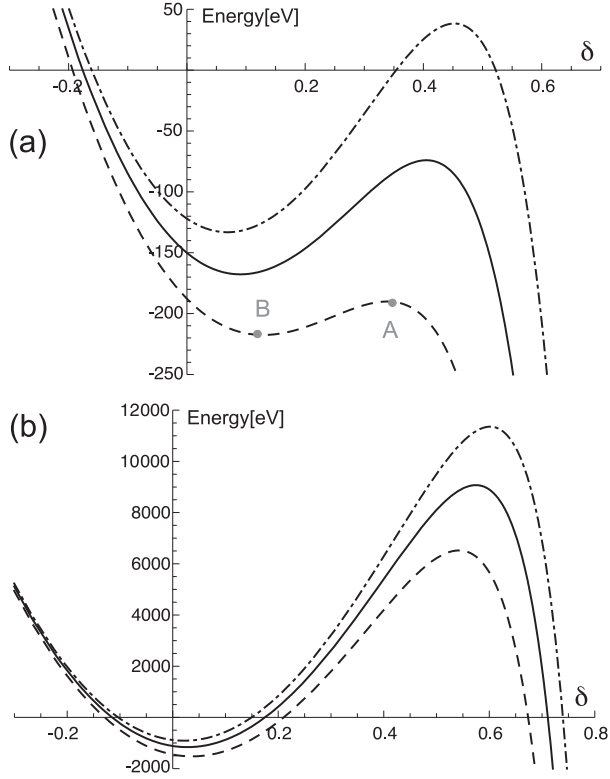}
\caption{\label{fig4}Potential profiles for $CC$ configuration (dashed line), $CA$ configuration (continuous line) and $AA$ configuration (dot-dashed 
line). In the (a) plot $x_0=100$ $nm$, $k=0.1$ $N/m$ and $A=2\cdot 10^{-10}$ $m^2$ letters A and B refer to Fig. \ref{fig5}(a); in the (b) plot $x_0=200$ $nm$, $k=0.5$ $N/m$ and $A=10^{-8}$ 
$m^2$.}
\end{figure}

Device instability is usually discussed in terms of a dimensionless parameter, like the ratio between the Casimir energy at the equilibrium distance and the elastic energy \cite{serry,barcenas,barta}. This can be done conveniently when the force has a simple inverse power dependence with the separation distance. The analysis of the bifurcation diagrams as a function of the parameter allows one to draw conclusions about changes in the 
critical separation with the material dielectric properties.
In our case the force as a function of the separation distance cannot be reproduced by a simple power law, so that we cannot adopt a single dimensionless parameter to study the device instability.\\
The behaviour of the device depends upon the spring constant $k$, the area $A$ of the plates and the unactuated distance $x_0$. The ratio $\Xi =A/k$ 
gives a measure of the relative importance of the vacuum and the elastic force, since they increase linearly with $A$ and $k$ respectively. According 
to the values attributed to these parameters one obtains different intervals of distances and energies of device operation at fixed $x_0$. To show the effects of 
the phase transition we give in Fig. \ref{fig4} plots of the potential energy of the device for fixed $k$ and $A$ at $x_0$ values of the order of 
hundred nanometers. 
This choice corresponds to distances where the relative difference of the force varies significantly (see Fig. \ref{fig3}). As expected, on passing 
from the $CC$ to the $AA$ configuration the weakening of the vacuum force shifts the stable equilibrium position towards smaller $\delta$ values and 
the maximum at larger $\delta$. The displacement is more pronounced for low $\Xi$ i.e. soft spring and small area (Fig. \ref{fig4}(a)), where it can 
be of the order of $80\%$ and more, than in the 
case of large $\Xi$ plotted in Fig. \ref{fig4}(b), where the increase of the travel range is of the order of $10\%$. The range of energies for 
which the device can be operated is modified in a similar way. These curves provide examples of the possibility of tuning in reversible way the 
device properties using the metal-insulator phase transition.
\begin{figure}
\centering
\includegraphics[width=8cm,angle=0]{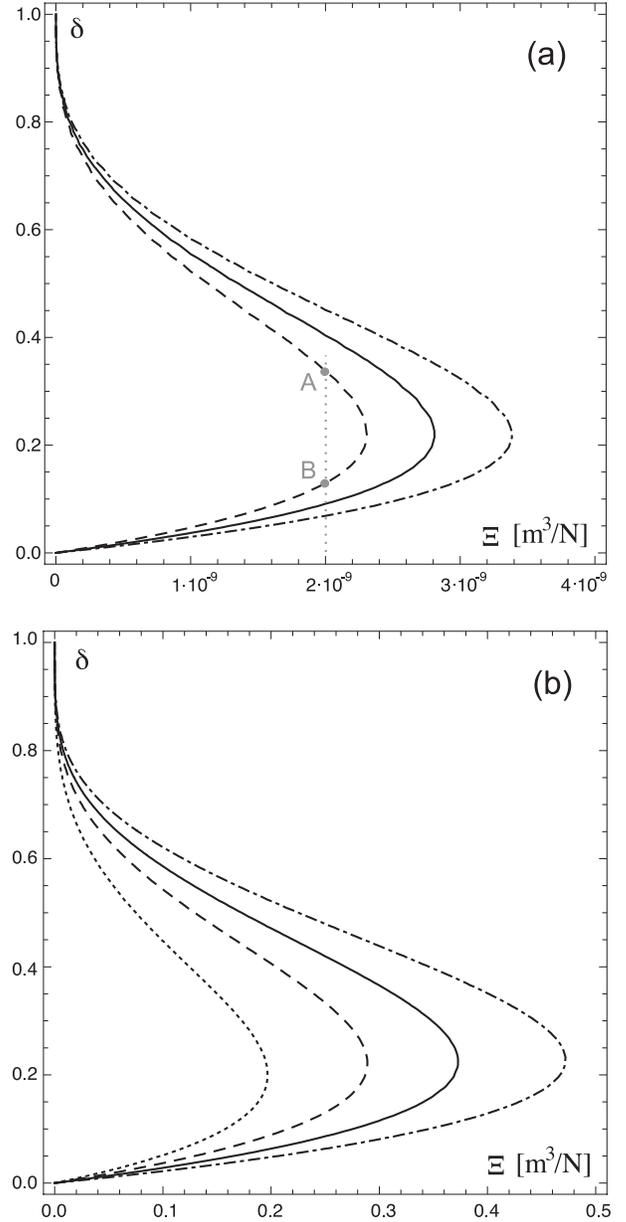}
\caption{\label{fig5}Bifurcation diagrams for $CC$ configuration (dashed line), $CA$ configuration (continuous line) and $AA$ configuration (dot-dashed line). In the (a) plot $x_0=100$ $nm$, the dotted line represents the $\Xi$ value used in Fig. \ref{fig4}(a). In the (b) plot $x_0=5$ $\mu m$, the dotted line represents the force calculated using the $T=0^{\circ}$ $K$  Casimir's force.}
\end{figure}
Fig. \ref{fig5} displays bifurcation diagrams of the device as a function of $\Xi$ for $x_0$ equal to $100$ $nm$ (Fig. \ref{fig5}(a)) and $5$ $\mu m$ 
(Fig. \ref{fig5}(b)) \cite{barcenas,pelesko,lin} for the $CC$, $CA$ and $AA$ configurations. The curves have similar shapes, with a lower branch 
before the fold that corresponds to stable solutions while the upper branch indicates the unstable states. The $\delta$ values which allow to operate 
the device at a given value $\Xi^{\star}$ are those which lie within the intersections of the line  $\Xi= \Xi^{\star}$ with the upper and the lower 
branch of the curves (points A and B in Fig. \ref{fig5}(a)). At the $\Xi$ value corresponding to the curve fold there is only one intersection, which 
gives the critical separation $\delta_0$. As expected, the crystalline-amorphous transition changes significantly the range of $\Xi$ parameters over 
which the device can work. Notice that $\delta_0$ varies between $0.21$ and $0.22$. The critical value is expected to change when the functional 
dependence of the vacuum force with the plate separation is modified: for the $d^{-4}$ dependence typical of the Casimir force $\delta_0=0.20$, while 
for the $d^{-3}$ behaviour, which is characteristic of the van der Waals interaction and of the large distance force at $T>0^{\circ}$ $K$, 
$\delta_0=0.25$ \cite{serry,barcenas}. In the case illustrated in Fig. \ref{fig5}(a), i.e. $100$ $nm$ distance, the critical value deviates significantly from the third 
inverse power behaviour. The curves of the $5$ $\mu m$ case show a similar behaviour. For the sake of comparison we have reported in Fig. \ref{fig5}(b) the curve 
corresponding to a pure Casimir interaction, which has been often adopted in similar device analysis. This force is appropriate to describe the case 
of metallic plate at large distances for $T=0^{\circ}$ $K$. Comparison with the results for our $CC$ configuration shows that the ranges of $\Xi$  
and $\delta$ values are considerably underestimated by this interaction. Notice that the vacuum force at such distance cannot yet be described by a 
third inverse power law, as expected from equation (\ref{ecco2}) at $T>0^{\circ}$ $K$. This is most clearly seen by the critical $\delta_0$ value 
which is smaller than $0.25$.\\
To better illustrate how the separation distance dependence of the vacuum force can affect the device performance, we plot in Fig. \ref{fig6} the 
critical value as a function of the unactuated distance. It is seen that $\delta_0$ varies between $0.25$ at very short (less than $10$ $nm$) and 
very large (more than $10$ $\mu m$) distances and a minimum slightly larger than $0.20$. The minimum occurs at different distances in the three 
configurations and it is found at smaller $x_0$ for the $AA$ case. It is clear from this curve that simulating device behaviour on the basis of 
simple inverse power law forces does not allow to get realistic results. 
\begin{figure}
\centering
\includegraphics[width=8cm,angle=0]{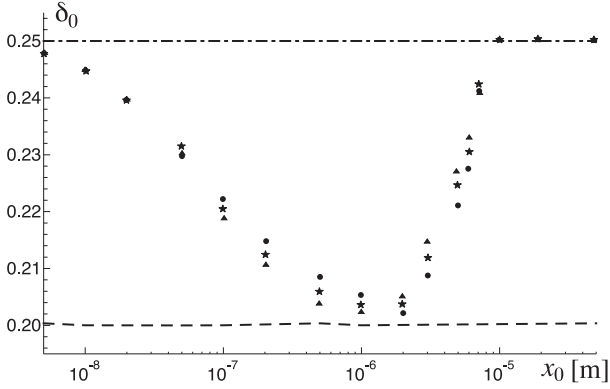}
\caption{\label{fig6}$\delta_0$ as a function of $x_0$. Dashed line corresponds to the $T=0^{\circ}$ $K$ Casimir's force, dot-dashed line is obtained in the small and large $d$ limits of equations (\ref{ecco1}) and (\ref{ecco2}). Dots represent the exact results for $F_{CC}$ (circles), $F_{AA}$ (triangles) and $F_{CA}$ (stars).}
\end{figure}
\section{Conclusions}
We have shown how the metal-insulator transition in germanium tellurides can be exploited in a simple device to modify its performance and the conditions under which it can be used. The choice of the appropriate values for the parameters entering into our model may depend upon a number of factor, like the film quality \cite{pirozhenko}, the possibility of controlling the kinetics of the phase transition, the mechanical properties of the components, the surface roughness \cite{Palasantzas2} etc. which have to be clarified in order to plan the realization of a specific device. On the theoretical side two aspects have to be further investigated. The first concerns the role of the film size in determining the change in the vacuum force: this can be carried out along the lines of previous studies \cite{zhou,barcenas} with the purpose of determining the minimum thickness that allows to detect significant vacuum force variations. The second has to do with the effects of the finite plate area, which in the metal case may be needed to avoid inconsistencies related to the Drude description of the intraband absorption \cite{geyer}.    
\acknowledgments
The authors want to acknowledge Stephan Kremers (RWTH Aachen University) for the material he shared with them.
AB is also grateful to Erio Tosatti (ICTP/SISSA Trieste) and Wojciech Welnic (LSI Ecole Polytechnique Palaiseau) for helpful discussions.
AB thanks CINECA Consorzio Interuniversitario for funding his PhD fellowship. 
\bibliographystyle{unsrt}
\bibliography{mybiblio}
\end{document}